\documentclass{ws-procs975x65}

\def\beq{\begin{equation}}
\def\eeq{\end{equation}}

\begin{document}

\title{ALFABURST: \\ A realtime fast radio burst monitor for the Arecibo
    telescope}
\author{Jayanth Chennamangalam$^{*1}$, Aris Karastergiou$^{1,2,3}$,
    David MacMahon$^4$, Wes Armour$^5$, Jeff Cobb$^4$, Duncan Lorimer$^{6,7}$,
    Kaustubh Rajwade$^6$, Andrew Siemion$^{4,8,9}$, Dan Werthimer$^4$,
    Christopher Williams$^1$}

\address{$^1$Astrophysics, University of Oxford,\\
Denys Wilkinson Building, Keble Road, Oxford OX1 3RH, UK\\
$^*$E-mail: jayanth@astro.ox.ac.uk\\
www-astro.physics.ox.ac.uk}

\address{$^2$Department of Physics and Electronics, Rhodes University,\\
PO Box 94, Grahamstown 6140, South Africa}

\address{$^3$Physics Department, University of the Western Cape,\\
Cape Town 7535, South Africa}

\address{$^4$Department of Astronomy, University of California Berkeley,\\ 
Berkeley, CA 94720, USA}

\address{$^5$Oxford e-Research Centre, University of Oxford,\\
Keble Road, Oxford OX1 3QG, UK}

\address{$^6$Department of Physics and Astronomy, West Virginia University,\\
PO Box 6315, Morgantown, WV 26506, USA}

\address{$^7$National Radio Astronomy Observatory,\\
PO Box 2, Green Bank, WV 24944, USA}

\address{$^8$ASTRON, PO Box 2, 7990 AA Dwingeloo, The Netherlands}

\address{$^9$Department of Astrophysics, Radboud University,\\
PO Box 9010, 6500 GL Nijmegen, The Netherlands}

\begin{abstract}
Fast radio bursts (FRBs) constitute an emerging class of fast radio transient
whose origin continues to be a mystery. Realizing the importance of increasing
coverage of the search parameter space, we have designed, built, and deployed a
realtime monitor for FRBs at the 305-m Arecibo radio telescope. Named
`ALFABURST', it is a commensal instrument that is triggered whenever the
1.4~GHz seven-beam Arecibo $L$-Band Feed Array (ALFA) receiver commences
operation. The ongoing commensal survey we are conducting using ALFABURST has
an instantaneous field of view of 0.02~sq.~deg. within the FWHM of the beams,
with the realtime software configurable to use up to 300 MHz of bandwidth. We
search for FRBs with dispersion measure up to 2560~cm$^{-3}$~pc and pulse
widths ranging from 0.128~ms to 16.384~ms. Commissioning observations performed
over the past few months have demonstrated the capability of the instrument in
detecting single pulses from known pulsars. In this paper, I describe the
instrument and the associated survey.
\end{abstract}


\bodymatter

\section{Introduction}

Fast radio bursts (FRBs) are non-repeating, millisecond-duration, broadband,
dispersed radio pulses that exhibit values of dispersion measure (DM) much
larger than what is expected due to the Galactic electron
density.\cite{lor07,tho13} This latter fact {\it prima facie\/} leads to the
inference that they are extragalactic in origin, with the millisecond-duration
pulse-widths implying compact object progenitors. Although a few
Galactic-origin hypotheses have been put forward (see, for example,
Ref.~\refcite{loe14}), most theories of FRBs present cosmological explanations,
including, for example, the collapse of supramassive neutron stars to black
holes,\cite{fal14} and Alfv\'{e}n wings around planets orbiting extragalactic
pulsars.\cite{mot14} Only a few FRBs are known, and all except one were
discovered using the Parkes telescope. The exception was discovered using the
Arecibo telescope.\cite{spi14} If FRBs are indeed extragalactic in
origin, they can be used as probes of the intergalactic environment, for
example, to measure the intergalactic baryon content, helping solve the
long-standing `missing baryon problem' in cosmology, wherein a discrepancy
exists between the expected and observed quantities of baryons in the
Universe.\cite{per92,mcq14} However, due to the large uncertainty in
localization of these bursts, a conclusive explanation of their origin is still
lacking. Recent efforts have focused on automated, realtime searches for FRBs
to try to increase the sample size,\cite{pet15,kar15} and, in case of
detection, perform signal characterization.\cite{pet15}

We have recently designed, developed, and deployed one such realtime FRB search
pipeline for the 305-m diameter Arecibo telescope. Named `ALFABURST', it is a
commensal data processing system that utilizes the 1.4~GHz seven-beam Arecibo
$L$-band Feed Array (ALFA) receiver. ALFA is Arecibo's workhorse survey
receiver, being used in the successful Pulsar ALFA (PALFA) survey,\cite{cor06}
that has so far resulted in the discovery of 145 pulsars, and one FRB.
ALFABURST piggybacks on the commensal data processing infrastructure developed
for the SERENDIP~VI system, the latest iteration of the SERENDIP series of SETI
spectrometers at Arecibo. It also partially re-uses ARTEMIS, the FRB search
code developed for the recently concluded ARTEMIS survey at the UK station of
the LOFAR telescope\cite{kar15}. A detailed description of the instrument is
given in Sec.~\ref{sec_sysdesc}.

For the commensal survey that we plan to conduct, the constituent observations
are triggered automatically whenever the primary observer enables the ALFA
receiver. Based on historical and expected usage of ALFA, we expect to observe
for 500--1000 hours over the next year. A detailed description of the survey
is give in Sec.~\ref{sec_surv}.

\section{System Description}\label{sec_sysdesc}

\begin{figure}
\begin{center}
    \includegraphics[width=\textwidth]{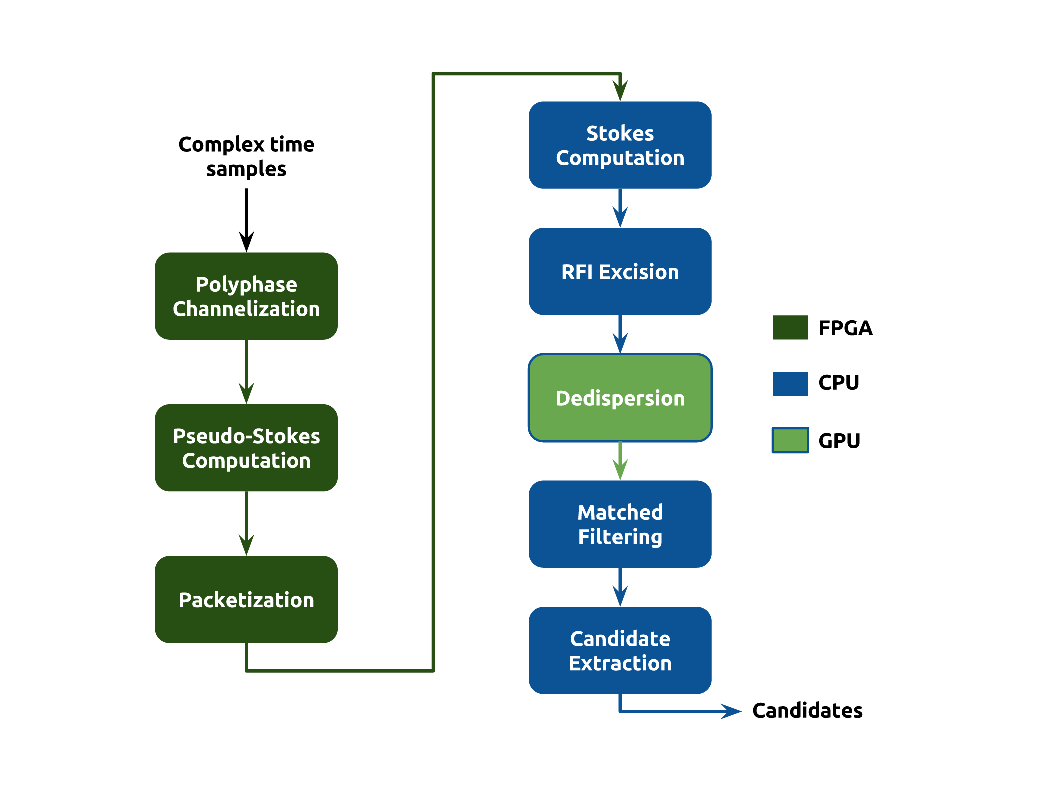}
    \caption{The signal processing steps in ALFABURST, as described in
        Sec.~\ref{sec_sysdesc}. The colour of the blocks represents the device
        the operation takes place in.}
    \label{fig_blockdiag}
\end{center}
\end{figure}

ALFABURST is a heterogeneous instrument, i.e., it combines two Field
Programmable Gate Array (FPGA) boards --- namely, the ROACH2, developed by the
CASPER collaboration\footnote{\url{http://casper.berkeley.edu/}} --- with a
high-performance computing (HPC) cluster equipped with gaming graphics
processing units (GPUs). The down-converted radio frequency signal from the
receiver is digitized using fast-sampling analog-to-digital converters at the
input of the FPGA boards and polyphase-channelized to 4096 channels. 16-bit
pseudo-Stokes values --- $XX^*$, $YY^*$, Re($XY^*$), and Im($XY^*$), where $X$
and $Y$ are the Fourier transforms of the two polarizations, and $X^*$ and
$Y^*$ are the respective complex conjugates --- are then computed. The data is
packetized and sent over 10-Gigabit Ethernet to the HPC cluster.

The HPC cluster is made up of five server-class computers, one of which is the
`head node', and the remaining four are `compute nodes'. The head node monitors
the receiver in use, and if ALFA is enabled, runs the data acquisition programs
on the compute nodes. The compute nodes are dual-socket, dual-GPU systems
equipped with NVIDIA GeForce GTX TITAN cards and RAID data disks. Each node,
barring the last, runs two instances of the software pipeline, each instance
handling a single ALFA beam. The last compute node handles the seventh beam.

The realtime pipeline first computes Stokes values from the pseudo-Stokes data,
and the total power is used for downstream processing. An adaptive-thresholding
RFI excision algorithm removes spurious noise and normalizes the data to have
zero mean and uniform RMS. This is followed by dedispersion, matched filtering,
and thresholding for candidate extraction. The dedispersion stage is based on
the GPU-based AstroAccelerate dedispersion transform algorithm.\cite{arm12}
Fig.~\ref{fig_blockdiag} shows a schematic of the signal processing operations
of the instrument.

\begin{figure}
\begin{center}
    \includegraphics[width=\textwidth]{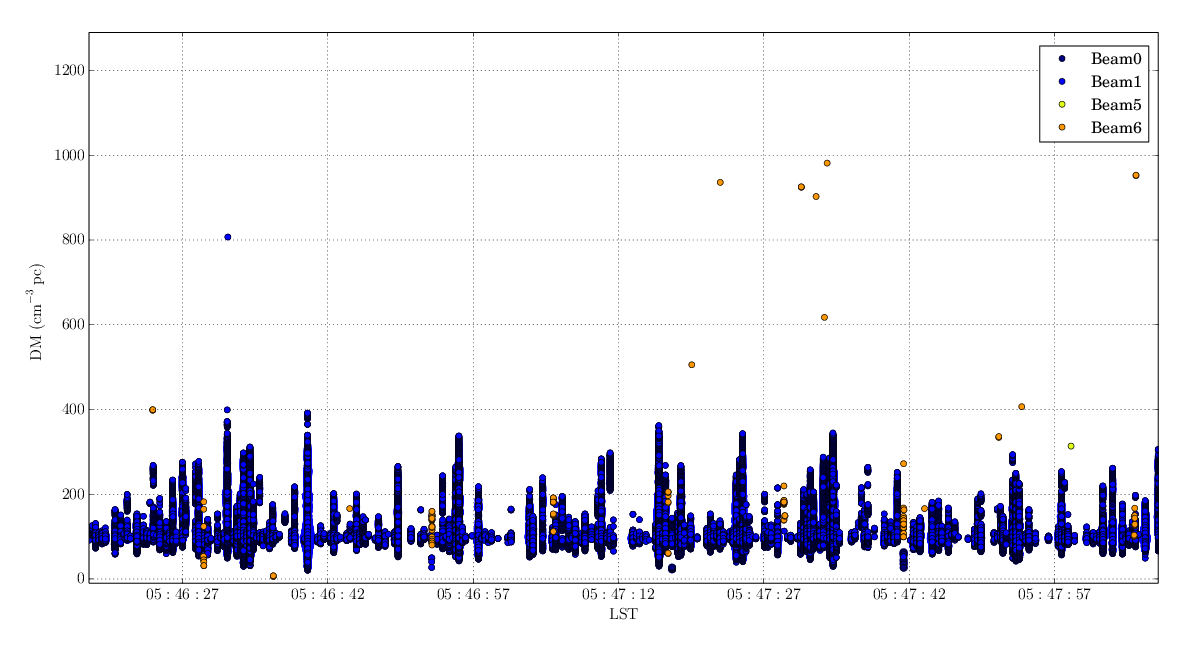}
    \caption{Results of a commissioning test of the system wherein we observed
    the pulsar B0611+22 using beam 1 of ALFA, with the pipeline processing the
    data in realtime. Each marker represents an event that crossed a 6-sigma
    threshold. The pulsar $S/N$ peaks around the expected DM of
    97~cm$^{-3}$~pc.}
    \label{fig_7beam}
\end{center}
\end{figure}

We performed commissioning tests of the system between March and July
2015 (for details, see Rajwade et al., these Proceedings). We observed a few
pulsars known for their strong single pulses using each of the seven ALFA
beams, and were able to blindly detect them using our pipeline.
Fig.~\ref{fig_7beam} shows detections from the pulsar B0611+22.

\section{Commensal Survey}\label{sec_surv}

For the commensal survey that we plan to conduct, observations are triggered
automatically whenever the primary observer enables the ALFA receiver. Based on
historical and expected usage of ALFA, we expect to observe for 500--1000 hours
over the next year. Using an event rate of
$3.1\times10^4$~sky$^{-1}$~day$^{-1}$ above 350~mJy that is based on the ALFA
FRB discovery of Ref.~\refcite{spi14}, an instantaneous field of view of
0.109~sq.~deg. that takes into account sensitive sidelobes, and the expected
usage of the receiver, we expect to detect 2--3 FRBs over the next year.

\section*{Acknowledgments}

We thank Dana Whitlow, Arun Venkataraman, Mike Nolan, Phil Perillat and Luis
Quintero at the Arecibo Observatory for their assistance during instrument
deployment and test observations. JC, AK and WA would like to thank the
Leverhulme Trust for supporting this work.

\end{document}